\documentstyle[12pt]{article}
\renewcommand{\theequation}{\arabic{section}.\arabic{equation}}
\newcounter{saveeqn}
\newcommand{\alpheqn}{\setcounter{saveeqn}{\value{equation}}%

\setcounter{equation}{0}\addtocounter{saveeqn}{1}%

\renewcommand{\theequation}{\arabic{section}.\arabic{saveeqn}.
\alph{equation}}}
\newcommand{\reseteqn}{\setcounter{equation}{\value{saveeqn}}%
\renewcommand{\theequation}{\arabic{section}.\arabic{equation}}}

\begin{document}

\begin{titlepage}\hfill HD-THEP-96-10
\vspace{1.5cm}
\begin{center}{\Large\bf Optimized $\delta $ - Expansion\\ for Lattice
U(1) and SU(2)\\ with Interpolating Continuum Action}\\\vspace{2cm}

{\bf Iring Bender and Dieter Gromes}\\\vspace{1cm}

Institut f\"ur Theoretische Physik der Universit\"at
Heidelberg\\Philosophenweg 16, D-69120 Heidelberg \\

E - mail: i.bender@thphys.uni-heidelberg.de\\
\hspace{9.0ex}d.gromes@thphys.uni-heidelberg.de        \end{center}
\vspace{2cm}
{\bf Abstract: } Embedding the lattice gauge theory into a continuum
theory allows to use the continuum action as trial action in the
variational calculation. Only originally divergent graphs contribute.
This leads to a very simple scheme which makes it possible to write
down explicit expressions for the plaquette energy $E$ for U(1) in
arbitrary space time dimension for the first three orders of the
expansion. For dimensions three and four one can even go up to fourth
order. This allows a rather thorough empirical investigation of the
convergence properties of the $\delta $-expansion, in particular near
the phase transition or the transition region, respectively. As
already found in previous work, the principle of minimal sensitivity
can be only applied for $\beta $ above a certain value, because
otherwise no extremum with respect to the variational parameter
exists. One can, however, extend the range of applicability down to
small $\beta $, by calculating instead of $E$ some power $E^\kappa $,
or by performing an appropriate Pad\'e transformation. We find
excellent agreement with the data for $\beta $ above the transition
region for the second and higher orders. Below the transition region
the agreement is rather poor in low orders, but quite impressive in
fourth order. For SU(2) we performed the calculation up to second
order. The agreement with the data is somewhat worse than in the
abelian case.

\vfill \centerline{April 1996}\end{titlepage}
\setcounter{equation}{0}\addtocounter{saveeqn}{1}%
\section{Introduction }

The ``optimized $\delta $ - expansion'',  also called ``linear
$\delta $ - expansion'', or, more appropriately, ``variational
perturbation theory'', is a powerful method which combines the merits
of perturbation theory with those of variational approaches. The
underlying idea is simple: The action $S$ is split into a
free part $\lambda S_0$ and an interacting part $S-\lambda S_0$.
Actually it is not necessary that $S_0$ describes a free action, but
only that all relevant quantities can be calculated
explicitely with $S_0$. The interacting part is multiplied by a factor
$\delta $ which serves as expansion parameter and is put equal to one
at the end. The exact result should be independent of the parameter
$\lambda $ while any approximation  will depend on it. The idea,
often called ``Stevenson's principle of minimal sensitivity''
\cite{stev}, is that the approximate solution should depend as
little as possible upon the parameter $\lambda $. This means that
$\lambda $ should be chosen such that the quantity to be calculated
has an extremum. In this way the result becomes non-perturbative
because $\lambda $ becomes a non-linear function of the coupling
constant. In every order of perturbation theory the parameter has to
be calculated again. There are already many successful applications of
this method in various fields of physics as well as rigorous
convergence proofs for simple cases. We refer, e.g. to the references
given in \cite{Gr}. In the present paper we concentrate on
applications on the lattice. Up to now, three different types of
actions $S_0$ have been used in this context.

In the first paper on the subject by Duncan and Moshe \cite{DM},
which, among other topics, treats the plaquette energy for U(1) and
$Z_2$ gauge theory in $d = 3$ space time dimensions, the action $S_0$
was chosen as a maximal tree of plaquettes. This is a set of
plaquettes which does not contain a closed surface, whereas the
addition of any further plaquette would lead to a closed surface. One
of the reasons for this choice is, of course, that all integrations
can be performed explicitely in this case. A maximal tree for $S_0$
was also used by Buckley and Jones \cite{BJ2} in their work on Z(2),
U(1), and SU(2) in four dimensions.

A second natural choice for $S_0$ is a quadratic action, typically the
sum of the squares of the plaquette angles. Such an action was used by
Duncan and Jones \cite{DJ} in their work on U(1) in $d = 4$ and by
Buckley and Jones \cite{BJ1} for SU(2).

A third choice, used by Akeyo and Jones \cite{AJ} for SU(2), as well
as for the mixed SU(2) - SO(3) model, is a single link action, i.e.
the sum of $\mbox{Tr}\;U_l$ over all links.

A maximal tree is a good approximation to the original action in the
strong coupling limit. A quadratic action or a single link action,
on the other hand, is a good approximation in the weak coupling limit.
The $\delta $ - expansion therefore behaves quite differently in the
two cases: In the first case one obtains a good description of the
Monte Carlo data for small $\beta $, roughly up to the phase
transition, or the transition region, respectively. In the second case
the same holds for large $\beta $ from the transition region up to
infinity.

The signal for the qualitative change in the behavior is the merging
of two extrema (with respect to $\lambda $) into a point of inflexion with
horizontal tangent and it's subsequent disappearance, when $\beta $ is
changed near the critical region. We will, however, see, that this
point of inflexion has no special relevance.

In the present paper we use a fourth ansatz for $S_0$ which has
several advantages compared to the previous ones (as well as a minor
drawback). We enlarge the degrees of freedom of the system by
embedding the lattice gauge theory into a continuum theory and use the
free Maxwell Lagrangian for the vector potential as our interpolating
action. The divergences which arise in the continuum are absorbed by
splitting off a divergent factor from the action. The consequence
thereof is that only originally divergent graphs, in which a photon
line starts and ends at one and the same link, survive. This leads to
a dramatic simplification. No graph has to be calculated explicitely,
the whole game essentially becomes a problem of counting certain
configurations of plaquettes. The advantages of the method are the
following:

\begin{itemize}

\item In any order of the calculation we obtain neither integrals, nor
infinite sums, nor special functions.

\item In any order $n$ of the expansion only a finite number of
configurations, consisting of $n+1$ connected plaquettes, has to be
considered.

\item In any order the result only contains polynomials in the
variational parameter $\lambda $ and powers of $e^{-\lambda /4}$.

\item The calculation can be easily performed for arbitrary dimension.

       \end{itemize}

These features allow calculations to a comparatively high order. In
the present paper we obtained explicit expressions for arbitrary
dimension $d\geq 3$ up to third order of the $\delta $- expansion. For
the cases $d = 3$ and $d = 4$, with the help of a computer program
which searches the relevant configurations of plaquettes, we can go to
fourth order, one order more than computed in \cite{DJ}.

We also mention here a drawback of the method:

\begin{itemize}

\item The coefficient in front of $1/\beta $ in the weak coupling
expansion is not correctly reproduced from the beginning, but it
converges to the correct value in higher orders.

\end{itemize}

Since our main interest is in the transition region, this drawback can
be easily tolerated. A priori it is anyhow impossible to tell which
type of trial action is best suited in this region.

In sect. 2 we explain our method for U(1) and give the relevant
general formulae. In sect. 3 we present the results up to order 4.
Here we also discuss possibilities to enlarge the region in $\beta $
where the principle of minimal sensitivity can be applied. In sect. 4
we extend the method to SU(2) and apply it up to second order.
Sect. 5 summarizes our conclusions. \newpage

\setcounter{equation}{0}\addtocounter{saveeqn}{1}%

\section{The method for U(1) }
We use the familiar formulation of U(1) on a $d$-dimensional lattice
with lattice constant $a$, described by the
partition function

\begin{equation} Z=\int _{-\pi }^{\pi }\cdots \int _{-\pi }^{\pi
}e^{-\beta S} \prod _l \frac{d\phi  _l} {2\pi },\end{equation}
with the action

\begin{equation} S = -\sum_{p'}\cos \Theta _{p'}.
\end{equation}
Here $l$ runs over the links, $p'$ over the plaquettes, $\Theta _{p'}$
is the sum of the four (oriented) angles $\phi _l$ living on the links
of the plaquette $p'$. The object we shall discuss is the average
plaquette energy E, i.e. the expectation value of $\cos \Theta_p$.

In the first step we want to extend the integrations over the angles
$\phi _l$ to the interval from $-\infty$ to $\infty$. This can be done
by using the following relation which holds for any periodic function
$f(\phi )$:

\begin{equation} \int _{-\pi }^\pi f(\phi ) \frac{d \phi } {2\pi } =
\lim_{\gamma \rightarrow 0} \sqrt{2 \gamma } \int_{-\infty}^\infty
e^{-\gamma \phi ^2} f(\phi ) \frac{d\phi } {2\pi }. \end{equation}
This is easily proven by splitting the rhs into integrals of length
$2\pi $, shifting the integration variable back into the interval from
$-\pi $ to $\pi $, and replacing the sum over the intervals by an
integral in the limit $\gamma \rightarrow 0$. Actually we will
see later that in any finite order of our expansion we can simply put
$\gamma = 0$ in the expectation value, which is a pleasant
simplification.

In the next step we enlarge the number of degrees of freedom
drastically, by introducing a vector potential $A_\mu $ defined in the
whole continuum. The connection with the link variables $\phi _l$
is, as usual,
\begin{equation} \phi _l = e\int _l A_\mu d x_\mu ,\end{equation}
whith $e^2 = 1/\beta $. Expectation values are not changed if we now
replace the ordinary integrations $\prod _l d\phi _l/(2\pi )$ by the
path integral ${\cal D}[A]$. The reason for this is easily understood:
The fields which are not sitting on the links appear neither in the
action nor in the plaquette energy. Therefore the corresponding
integrations factorize both in the numerator and in the denominator
and cancel. The same happens for the fields which sit on the links but
are transversal to them. Finally, for the longitudinal fields on the
links, one may go over to new variables by performing a linear
transformation with constant coefficients in such a way that one of
the variables becomes the integral $\phi _l$ in (2.4). The
integrations over the remaining other variables, as well as the
constant Jacobian, cancel again and we are left with the original
expectation value.

We can thus write the expectation value of the plaquette energy as an
expectation value in the continuum theory:

\begin{equation} E =\lim_{\gamma \rightarrow 0} \frac{1}{N_\gamma }
\int E_p \;\exp[\beta \sum_{p'}E_{p'}] \;\exp[-\gamma e^2\sum_l(\int_l
A_\mu d x_\mu )^2 ]{\cal D}A, \end{equation}
with
\begin{equation} E_p = \cos e\oint_p A_\mu  d x_\mu .
\end{equation}
The normalization constant $N_\gamma $ is, of course, obtained by
replacing $ E_p$ by 1 in the numerator of (2.5).

We are now in the position to apply the optimized $\delta $-expansion
by introducing the free continuum Lagrangian. All expressions in (2.5)
are gauge invariant in the limit $\gamma \rightarrow 0$, so, for
simplicity, we will use the Feynman gauge which has the advantage that
all graphs connecting orthogonal links vanish from the beginning.
While, on the lattice, everything is finite, we will immediately
obtain divergences in the continuum from the singularity of the
propagator at zero distance. At first sight this looks like an
additional complication. It can, however, easily be overcome by
splitting off an appropriate divergent constant from the free action.
In fact, this leads to a drastic simplification, because only the
divergent terms of ordinary perturbation theory survive. This, in
turn, will have the consequence that in any order only finite
connected sets of plaquettes are involved.

As free interpolating action we choose

\begin{equation} S_0 = -\frac{c} {\beta \lambda }\int A_\mu \Box A_\mu
d^dx. \end{equation}
Here $\lambda $ is the variational parameter. The constant $c$ is a
positive parameter which is divergent for $d\geq 3$. It is formally
defined by

\begin{equation} c = -\int_{0}^a\int_{0}^a D(t-t')dt\; dt'
\end{equation}
with $D$ the Green function of the d'Alembert operator, e.g. in four
dimensions $D(x)=-1/(4\pi ^2 x^2)$. Of course, one could easily use
some regularization which would lead to a large but finite $c$, and
later on perform the limit. Because the whole procedure is, however,
very transparent, this intermediate step can be skipped.

After introducing the free continuum action, the plaquette expectation
value in the optimized $\delta $-expansion reads

\begin{equation} E = \frac{1} {N(\delta )} \int E_p
e^{-S_0} e^{\delta (S_0 +\beta \sum_{p'}E_{p'})}{\cal D}A.
\end{equation}
We have simplified the expression by taking the limit $\gamma
\rightarrow 0$ in (2.5) which is now allowed in any finite order of
the $\delta $-expansion.

The problem has thus become a continuum problem of calculating
expectation values of products of plaquettes. The calculations for
higher orders are greatly simplified by a simple trick, essentially
already used in \cite{DM}. One should {\em not} expand the expression
(2.9) with respect to $\delta $ as it stands, because this would
introduce all the mixing terms between $S$ and $S_0$. Things become
much simpler if one leaves the term $(1-\delta )S_0$ together and
performs the substitution

\begin{equation} A_\mu = \sqrt{ \frac{\beta \lambda} {2c(1-\delta
)}}\;A'_\mu. \end{equation}
This brings the free action into the usual form $(1-\delta )S_0
\rightarrow S'_0 = -(1/2) \int A'_\mu \Box A'_\mu d^dx $. In $E_p$ and
$E_{p'}$ as defined in (2.6) one has to make the replacement

\begin{equation} e\rightarrow e' = \sqrt{\frac{\tilde{ \lambda }}
{2c}}, \end{equation}
where, for convenience, we have introduced the abbreviation

\begin{equation} \tilde{\lambda } \equiv \lambda /(1-\delta ).
\end{equation}
The quantity $\beta $ stays as it was before.
So we end up with the comparatively simple expression

\begin{equation} E=\int E_p e^{-S'_0} e^{\delta \beta \sum_{p'}
E_{p'}} {\cal D}A'/\int e^{-S'_0} e^{\delta \beta \sum_{p'}E_{p'}}
{\cal D}A' \end{equation}
in which now $e$ and $A_\mu $ are to be replaced by $e'$ and $A'_\mu
$ in $E_p$ and $E_{p'}$. Note, that besides the explicit $\delta
$-dependence of this expression, there is also an implicit $\delta
$-dependence contained in $e'$ which has to be considered.

Let us first look for the expectation value of the $1\times 1$ Wilson
loop with respect to the free continuum action. There are two types of
graphs: In the first type the propagator connects two different
parallel lines of the loop and is finite. The coupling constant $e'^2$
multiplying the propagator vanishes due to the constant $c = + \infty
$ in the denominator. Therefore the product is zero, i.e. the
exponential becomes equal to 1. We therefore only get contributions
from the self energy graphs where the propagator connects points on
the same link. For the four links of the plaquette this gives

\begin{equation} \exp\{\frac{4}{2}e'^2 \int_{0}^a\int_{0}^a D(t-
t')dt\; dt'\} = \exp (-\tilde{\lambda }) = \exp (-\lambda /(1-\delta
)). \end{equation}
Our choice for the divergent constant $c$ in (2.8) becomes clear from
this. In any order of the $\delta $-expansion the $\beta ^0$ term is
immediately obtained  by expanding (2.14) to the desired order in
$\delta $.

To obtain the complete result we have to expand (2.13). After
symmetrization in the summation variables $p_k$ one obtains a
series of the form

\begin{equation} E^{(n)} = \sum_{\nu =0}^n \frac{\eta _\nu
^{[n]} } {\nu !} \delta ^\nu \beta ^\nu \end{equation}
with

\begin{eqnarray} \eta _0 & = & <E_p>\nonumber\\
      \eta _1 & = & \sum_{p_1}[<E_p E_{p_1}> - <E_p><E_{p_1}>]\\
      \eta _2 & = & \sum_{p_1,p_2}[<E_p E_{p_1}E_{p_2}>
      -<E_p>< E_{p_1}E_{p_2}> - <E_{p_1}><E_p E_{p_2}>\nonumber\\
 &   &     - <E_{p_2}><E_p E_{p_1}> + 2 <E_p>< E_{p_1}><E_{p_2}>]
      \nonumber\\
 \cdots  & = & \cdots\cdots\cdots   \nonumber\\
 \eta _n & = & \sum_{p_1,\cdots ,p_n}[<E_pE_{p_1}\cdots
 E_{p_n}> - \mbox{\quad factorized contributions].\quad}\nonumber
\end{eqnarray}
Here $ <\cdots >$ denotes the normalized expectation value.
The coefficients $\eta _\nu ^{[n]}$ in (2.15) are defined by expanding
each $\eta _\nu (\tilde{\lambda })$ with respect to the $\delta $
contained in $\tilde{\lambda }$ up to order $n - \nu $. This means
that in total one expands up to order $n$ of the $\delta $-expansion.
Finally one has to put $\delta =1$.

The calculation of the expectation values proceeds along the following
scheme. Consider, e.g. the term $<E_pE_{p_1}\cdots E_{p_n}>$. Write
the cosines in the $E_{p_k}$ as exponentials, $\cos \Theta_k =
(1/2)\sum_{j_k=\pm 1}e^{ij_k \Theta_k}$, the cosine in $E_p$ can be
simply replaced by $e^{i \Theta}$. In this way one obtains a sum of
$2^n$ terms with a factor $2^{-n}$ in front. The expectation value
above is then evaluated with the use of the formula

\begin{equation}\frac{\int e^{-S'_0} e^{\int J'_\mu (x)A'_\mu (x)d^dx}
{\cal D}A'} { \int e^{-S'_0} {\cal D}A'} = \exp\{-(1/2)\int J'_\mu
(x)D(x-x')J'_\mu (x')d^dx\; d^dx'\}. \end{equation}
In our case all currents $J'_\mu $ are localized on the links and the
$d$-dimensional integrals above become one dimensional. The current on
a link has the form

\begin{equation} J_{link} = ie'\delta ^{(d-1)}(link)\sum_k
j_k^{(link)},\end{equation}
where $\delta ^{(d-1)}(link)$ is the $(d-1)$-dimensional $\delta
$-function with support on the link while the sum runs over all
plaquettes $p_k$ which share the considered link (put $p=p_0, j_0 = 1$
in this context). The further calculation is greatly simplified by the
fact that we never need any mixing terms between different links,
because these involve a finite propagator, so the exponent becomes
zero due to the factor $c$ in the denominator of $e'^2$. For the
singular diagonal contribution of a single link, on the other hand,
the divergent constant $c$ cancels and we end up with $-(\sum_k
j_k^{(link)})^2 \tilde{\lambda }/4$ in the exponent. Finally, we thus
obtain the generic formula

\begin{equation} <E_pE_{p_1}\cdots E_{p_n}> =\frac{1}{2^n}
\sum_{j_{k_1},\cdots ,j_{k_n}=\pm 1} \exp\{ -\sum_{links}(\sum_k
j_k^{(link)})^2 \tilde{\lambda }/4\}. \end{equation}
This allows all expressions in (2.16) to be evaluated in a simple way.

An enormous simplification arises through the fact that only connected
configurations of plaquettes need to be considered, where two
plaquettes are called connected if they share a common link (or are
identical). The reason is simple. If we have a configuration of two
sets of plaquettes which are disconnected from each other, there are,
as shown above, no contributions where the propagator connects the two
sets. The contributions thus factorize and are therefore canceled by
the factorized terms in (2.16). Therefore the expansion is local in
the sense that in any order $n$ there is only a finite number of
plaquettes, coming from the expansion of the exponent of the lattice
action, which needs to be considered. These plaquettes make up a
connected set together with the plaquette $p$. Therefore we simply
obtain a sum of expressions which only contain polynomials in $\lambda
$ (from the expansion of $\tilde{ \lambda }$) times powers of
$\exp(-\lambda /4)$. So, in any order we neither get integrations, nor
infinite sums, nor special functions! This simple structure allows the
calculation of comparatively high orders which would, e.g., become
prohibitively complicated in any approach working with lattice
propagators.

\setcounter{equation}{0}\addtocounter{saveeqn}{1}%

\section{Results for U(1)}
For the first three orders the plaquette configurations which
contribute in the sum and their contributions can be written down
explicitely. Only some modest computer help was used just for
convenience. In the following we will give the formulae for general
dimension $d$ but first discuss only $d = 4$. Other dimensions are
briefly treated at the end. \\[1ex]

{\bf  Order 1 and generalities}\\[1ex]

There are only two types of configurations in the sum over $p_1$ which
contribute (see fig. 1a). In the first type one has $p_1 = p$, its
contribution to $\eta _1$, according to the foregoing considerations,
is $<E_p^2> - <E_p>^2 \; = (1/2)(1+e^{-4\lambda } -2e^{-2\lambda}) $.
The second type consists of all plaquettes $p_1$ which share just one
link with $p$. Their number is $4(2d-3)$, where the factor 4 is, of
course, due to the four links of $p$, while the second factor counts
the possible orientations of $p_1$. All these plaquettes give the same
contribution $<E_pE_{p_1}> - <E_p><E_{p_1}> \; = (1 /2) (e^{-3\lambda
/2}+e^{-5\lambda /2}-2e^{-2\lambda })$. In this way one ends up with
the following result. \alpheqn

\begin{eqnarray} E^{(1)}(\beta, \lambda ) & = & \eta _0^{[1]} + \delta
\beta \eta _1^{[1]} = (1-\delta \lambda ) e^{-\lambda} +\delta \beta
\eta _1(\lambda ) \mbox{\quad with\quad}\\ \eta _1(\lambda ) & = &
\frac{1}{2 }[1+e^{-4\lambda } -2e^{-2\lambda} +4(2d-3)(e^{-3\lambda
/2}+e^{-5\lambda /2}-2e^{-2\lambda })].\end{eqnarray}\reseteqn
At the end, $\delta $ has, of course, to be set equal to 1. According
to the principle of minimal sensitivity we have to look for the
extrema with respect to $\lambda $. There is always a local maximum at
$\lambda = \infty $, corresponding to the limit, where we do not
introduce an interpolating continuum action at all, i.e. to the
ordinary strong coupling expansion. Choosing this extremum would
obviously lead to the expected result $E^{(1)} = \beta /2$. For large
$\beta $, on the other hand, there is always a minimum at small
$\lambda $. This is found by expanding $E^{(1)}(\beta ,\lambda ) = 1
-2\lambda +(3/2)\lambda ^2 +\beta (d+1/2)\lambda ^2+O(\lambda ^3)$.
The minimum is at $\lambda =1/(d+1/2)\beta +O(1/\beta ^2)$ and gives
$E^{(1)}=1-1/(d+1/2)\beta +O(1/\beta ^2)$. Obviously the qualitative
behavior in the weak coupling limit is correct, but the factor $1/d$
in front of $\beta $ is replaced by $1/(d+1/2)$ which means that it is
too small by 11 \% compared to the correct factor. The reason is, that
our continuum action is not an as appropriate approximation in the
weak coupling limit as, e.g. the quadratic action used in \cite{DJ},
\cite{BJ1}. We will see how this factor converges towards the correct
one in higher orders. Since our main interest is in the region of the
phase transition, the fact that we do not reproduce the correct weak
coupling limit in first order is only a minor drawback. In this
context one should also mention the merit of the variational method
that it anticipates to a large extend the higher order coefficients.
In our case, although it gives - 2/9 for the leading coefficient
instead of - 1/4, as just discussed, it gives, e.g. a second order
coefficient of $-2/81$ for $d = 4$. This is $79\%$ of the correct
second order term $-1/32$.

The full structure of the extrema of (3.1) is easily discussed and
essentially independent of the dimension $d$. For large $\beta $ there
are 3 finite extrema (in addition to the one at infinity), the
one with the smallest $\lambda $ is the minimum just discussed and
has to be chosen. If $\beta $ is decreased, this minimum and the
neighboring maximum merge into a turning point with horizontal
tangent, i.e. a point of inflexion. In the sense of catastrophe theory
one has a fold catastrophe there. The value where this happens is
easily found by solving the simultaneous equations $\partial
E/\partial \lambda =\partial^2 E/\partial \lambda ^2 = 0$. The
solution is $\beta _{pi} = 0.9674$. In fig. 2a we show our results for
order 1 to 4 together with the Monte Carlo data of Caldi \cite{Cal}.
We followed the minimum with the smallest $\lambda $ when coming from
large $\beta $ up to the point of inflexion at $\beta _{pi}$ where it
disappears.

The appearance or disappearance of extrema might be interpreted as a
signal for the existence of a phase transition. In fact, the value of
$\beta _{pi}$ found in this first order calculation is too small only
by 4\% compared to the value $\beta _c = 1.0081 \pm 0.0067$ given in
\cite{Cal}. But, on the other hand, one also finds a turning
point for $d=3$ where there is no phase transition.

There is a simple argument which shows that the point of inflexion has
no direct relation to the position of the phase transition. If this
were the case one should essentially obtain the same point of
inflexion if, instead of $E$, ones calculates some function of $E$,
say a power $E^\kappa $. In the spirit of the $\delta $-expansions one
has to calculate $E^\kappa $ from (3.1), expand to first order in
$\delta $, and finally put $\delta =1$. One finds that the position of
the point of inflexion depends drastically on $\kappa $. For $\kappa
=5$, e.g. one gets $\beta _{pi}=1.6089$. If $\kappa $ is decreased,
$\beta _{pi}$ decreases monotonically. At $\kappa _0=0.2781$ one
reaches $\beta _{pi} = 0.8252$, while for even smaller $\kappa $ there
is no point of inflexion at all, while the extrema persist! If one
chooses some $\kappa \leq \kappa _0$ one can therefore follow the
minimum down to small $\beta $ and calculate $E^\kappa $. From this
one may finally obtain $E$.

Before we investigate, whether one can obtain reasonable results by
this simple trick even below the transition region, let us first
mention, that the above considerations can serve as an excellent test
for the stability of the approach. Notwithstanding the fact that the
point of inflexion moves with the power $\kappa $, the value of $E$
finally obtained, should be essentially independent of $\kappa $
within a reasonable range. This is indeed the case to an impressive
accuracy. For illustration we choose the arbitrary value $\beta = 1.2$
and vary $\kappa $ in the range from $\kappa = -1$ to $\kappa \approx
2.3$ where $\beta _{pi}$ becomes equal to 1.2. The value for $E$ then
only moves from 0.7903 to 0.7868! The test becomes a bit worse if we
apply it to values of $\beta $ below $\beta _{pi}$, say $\beta =0.9$,
which can only be reached by the above trick for $\kappa $ small
enough. Varying $\kappa $ from -1 to 0.6, one finds that $E$ moves
from 0.7008 to 0.6917. These values still lie considerably above the
MC data but convergence to the correct values in higher orders can be
expected.

One may also perform a Pad\'e transformation with respect to $\delta $
before applying the principle of minimal sensitivity. This was done by
Duncan and Moshe \cite{DM} for the second order, in order to obtain an
extremum. In the first order discussed at the moment, the (0,1) Pad\'e
approximation has the interesting property that there is no turning
point where the extrema disappear. Therefore one can follow the
minimum over the whole range of $\beta $. This again shows that the
point of inflexion has no direct relevance. It demonstrates, however,
once more the impressive stability of the method for the values of
$\beta $ above the transition region. As seen in fig. 2b, for small
$\beta $ the power curve lies below the Pad\'e curve and closer to the
data, for $\beta >1$ both curves as well as the original first order
curve differ by less than 0.003. \\[1ex]

{\bf Order 2 }\\[1ex]

In second order there are 5 types of connected plaquette
configurations which contribute (fig. 1b). The first two of them are
identical with the ones of the first order, with one plaquette
occupied twice. To count the number of equivalent configurations
belonging to every type, one has to note that one of the plaquettes is
always identical to the fixed plaquette $p$, the other two have to be
arranged in all possible ways.

The discussion proceeds along the same lines as before. Contrary to
the first order there is now {\em no} relevant minimum near $\lambda
=0$ in the weak coupling limit, so that the principle of minimal
sensitivity cannot be directly applied. This is a well known feature
in simple models as the anharmonic oscillator in zero and one
dimension \cite{Osc}, where all even orders show the same behavior.
In our case, there is, however, a relevant minimum for values of
$\beta $ around $\beta \approx 1$, which merges with a maximum at
$\beta _{pi} =1.0168$. To this belongs the value $E_{pi} = 0.6783$ in
good agreement with the MC data. If one follows the minimum from the
point of inflexion to increasing $\beta $ one finds, however, that the
curve no longer follows the data. There is no extremum which
corresponds to the physical value. This is, of course, nothing but the
afore mentioned absence of a reasonable weak coupling result in even
orders.

To extract more useful information from the second order calculation
we apply the (1,1) Pad\'e transformation with respect to $\delta $ as
done by Duncan and Moshe \cite{DM} (the (0,2) transformation gives no
extrema at all). One finds a relevant minimum for all $\beta $ above
$\beta _{pi} = 1.0486$. The results are again presented in fig.
2a. They show considerable improvement compared to the first order and
already a very close agreement with the MC data. \\[1ex]

{\bf  Order 3 }\\[1ex]

In order three there are 16 types of connected configurations, 7 of
them are lower order configurations with multiply occupied plaquettes.
The correct counting of the number of equivalent members of one type
becomes delicate for some cases but is still feasible. We refer to
fig. 1c for details.

There is now again a reasonable weak coupling limit. For large $\beta
$ there is an extremum at $\lambda \approx 0.2234/\beta $ for $d =
4$ which leads to $E \approx 1-0.2417/\beta $. The error in
the coefficient of $1/\beta $ has become smaller by a factor of 3.4
compared to the first order. If one decreases $\beta $ and follows the
minimum, the latter disappears at $\beta _{pi} = 1.2187$. But at
some larger $\lambda $ there is another minimum. One may switch to
this with practically no change in the plaquette energy, and go
further down to $\beta _{pi} = 1.0625$ where this minimum also
disappears. The result shows only a minor change compared to the
second order and there is thus again excellent agreement with the
data.

As in the first order one can now again enlarge the region of
applicability by the power trick or the Pad\'e transformation. Only
the (0,3) transformation works, the (2,1) and the (1,2) transformation
show no extrema below the transition region. The results are shown in
fig. 2b. Again the power curve lies below the Pad\'e curve and both of
them are much closer to the data below the phase transition than in
first order. The discrepancy is, however still sizeable in this
region. For $\beta > 1.1$ both curves practically agree with the
ordinary third order curve. \\[1ex]

{\bf  Order 4 }\\[1ex]

The simplicity of our approach permits to go up to fourth order with
reasonable effort for $d = 3$ and $d = 4$. To do this we wrote a
computer program in Mathematica. It searches all possibilities for
connected plaquettes in a certain order. Some configurations which are
obtained from others by permutations of $p_1,\cdots,p_n$ are not found
in this way, while others which involve multiply coccupied plaquettes
are obtained several times. This is taken into account by applying the
appropriate factors. Finally the program calculates the contribution
for each configuration and adds up everything. This program also
served as a check for the lower order calculations.

As in the second order we apply a Pad\'e transformation. The only
useful one turned out to be the (2,2) diagonal transformation. There
are two intervals in $\beta $ where the use of the relevant extrema
gives good, respectively excellent, agreement with the data, as seen
in fig. 2a. The first one is {\em below} the transition region and
impressively reproduces the steep increase of the plaquette energy in
this region. For $\beta > 1.08$, however, the curve lies above the
data. There is a second interval, $1.4 < \beta < \infty $, where one
has perfect agreement with the data. For $\beta < 1.4$, however, the
curve lies again too high.\\[1ex]

{\bf  Other dimensions}\\[1ex]

The discussion proceeds as before, therefore we can be very brief
here. In first order the points where the minima disappear are at
$\beta _{pi} = 1.1937$ for $d = 3$ and at $\beta _{pi} = 0.8099$ for
$d = 5$. The results are shown in figs. 3a and 4a, the curves obtained
with the power trick and the (0,1) Pad\'e transformation in figs. 3b
and 4b. In second order the limiting values for the (1,1) Pad\'e
transformations are $\beta _{pi}= 1.4495$ for $d = 3$ and $\beta _{pi}
= 0.8090$ for $d = 5$. In third order one finds the following common
feature for large $\beta $: For increasing $\lambda $ there are two
minima and two maxima at finite $\lambda $, the extremum at lowest
$\lambda $ is the relevant minimum. If one decreases $\beta $, there
is a qualitative difference for different dimensions. In the case $d =
3$, the first minimum and maximum merge into a point of inflexion at
some $\beta _{pi}$. The second pair merges already for slightly larger
$\beta $, but this is of no importance, because these extrema are not
relevant. For $d = 4$ and $d = 5$, on the other hand, the first pair
merges at a larger $\beta $ than the second one, so one has to jump to
the second minimum for a short interval till this also disappears. For
$d\geq$ 6, finally, the irrelevant interior pair of extrema merges
first, for still smaller $\beta $ the outer pair merges into a point
of inflexion. This means that for $d = 3$, as well as for $d\geq$ 6,
one can follow the minimum down to the critical value, while for $d =
4$ and $ d = 5$ one has to jump to the other minimum near the phase
transition. Clearly the results are much better in five dimensions
than in three. This is due to the above mentioned error in the
coefficient of the weak coupling expansion which becomes smaller for
larger dimension. \newpage

\setcounter{equation}{0}\addtocounter{saveeqn}{1}%

\section{SU(2) }
We use the notation of Creutz \cite{Cr}, Lautrup and Nauenberg
\cite{LN}, and Buckley and Jones \cite{BJ1}, with $(\beta /2)
\sum_{p'} \mbox{Tr} U_{p'}$ in the exponent, and the plaquette energy
defined as the expectation value of $(1/2) \mbox{Tr} U_p$. In the non
abelian case it becomes crucial to choose an appropriate
parametrization for the unitary matrices $U_l$ on the links. A very
convenient parametrization with a simple behavior in the weak
coupling limit is the one proposed by Buckley and Jones \cite{BJ1}

\begin{equation} U=e^{i\sigma _1\varphi }e^{i\sigma _2\vartheta }
e^{i\sigma _3\psi }. \end{equation}
As parameter space one may use the region

\begin{equation} -\pi <\varphi,\psi <\pi ,\quad -\pi /4
<\vartheta<\pi /4. \end{equation}
Actually the group manifold is covered twice by this choice, but in
this way we will already get periodicity in $\varphi  $ and $\psi  $
immediately. The Haar measure, in a normalization convenient for us,
reads

\begin{equation} H(\psi ,\vartheta ,\varphi ) = \frac{\pi }{2}
\cos(2\vartheta).    \end{equation}
An efficient technique for the further procedure, which was also
extensively used in \cite{BJ1}, is the splitting of the matrix
exponentials in (4.1) into sums of ordinary exponentials times
projection operators, in general

\begin{equation} e^{i\sigma _k\alpha } = \sum_{s=\pm 1} e^{i s
\alpha } P_s^{(k)},\mbox{\quad with\quad }P_s^{(k)} = \frac{1}{2 }
(1+s\sigma _k).\end{equation}
From this it is immediately clear that all traces are periodic
functions of the three link angles $\varphi  _l,\vartheta _l,\psi  _l$
with period $2\pi $. So, for $\varphi  $ and $\psi  $ we may use the
same procedure as in the U(1) case, in order to extend the
integrations from $- \infty $ to $ \infty $. For the $\vartheta$
integration, on the other hand, the presence of the Haar measure and
the limited integration region from $-\pi /4$ to $\pi /4$ enforce a
special procedure. We have to continue the Haar measure periodically
into the full interval from $-\pi $ to $\pi $ by expanding it into a
Fourier series. This results in

\begin{equation} \frac{\pi }{2 }\cos(2\vartheta)_{periodic} =
\sum_{\nu =- \infty }^ \infty \frac{(-1)^\nu e^{4i\nu
\vartheta}}{1-4\nu ^2}. \end{equation}
The unitary matrices $U$ in (4.1) can be decomposed into a linear
combination of $1,\sigma _k$. It is then easily seen that they are
invariant under the substitution $\vartheta \rightarrow \pi /2 -
\vartheta$, if, simultaneously, one substitutes $\varphi \rightarrow
\varphi +\pi /2,\psi \rightarrow \psi +\pi /2$. The integral is
invariant under the latter substitutions due to the periodicity in
$\varphi $ and $\psi $. Together with similar relations, the
$\vartheta$-integration can thus also be extended to the interval from
$-\pi $ to $\pi $ and subsequently from $- \infty $ to $ \infty $ if
we use the periodic Haar measure in (4.5).

Next we introduce again the continuum fields $A_\mu ^{(a)}$, the
longitudinal components on the links are connnected to the link angles
by

\begin{equation} \varphi =\frac{g}{2 }\int {\bf A}^{(1)}d{\bf x},
\quad \vartheta =\frac{g}{2 }\int {\bf A}^{(2)}d{\bf x}, \quad \psi
=\frac{g}{2 }\int {\bf A}^{(3)}d{\bf x},\mbox{\quad with\quad
}g^2=4/\beta . \end{equation}
Contrary to the abelian U(1) case the procedure is now no longer gauge
invariant, because we introduce an ordinary exponential, not a path
ordered exponential, along the link. We have, of course, the freedom
to to this.

Next we introduce the free interpolating action as in (2.7), with the
sole difference that we have to sum over the three SU(2) indices $a$
in the potentials $A_\mu ^{(a)}$. The situation is now quite similar
to the abelian case. The expansion for $E^{(n)}$ looks as in (2.15),
(2.16) where now $E_p = (1/2) \mbox{Tr} U_p$. Again only connected
configurations of plaquettes contribute in the expansion.

The plaquette actions, when evaluated by using (4.4), contain 10 (not
12) projection operators, because two of the unitary matrices enter as
adjoints, and neighboring identical $\sigma $ - matrices can be
combined. The traces are therefore 10 fold sums with $2^{10}$ terms.
The plaquette angles appear in exponentials only. The path integrations
over $\varphi $ and $\psi $ (more precisely, those over $ A_\mu
^{(1)}$ and $ A_\mu ^{(3)})$ can be performed as before. In the
$\vartheta $ - integrations one has to consider in addition the
periodic continuation of the Haar measure (4.5), which also only
contains exponentials. In this way the following functions, defined by
infinite sums, arise:\alpheqn

\begin{eqnarray} f_m(\lambda ) & = &\sum_{\nu = - \infty }^ \infty
\frac{(-1)^\nu e^{-(4\nu +m)^2\lambda /4}}{1-4\nu ^2 }\\
h_m(\lambda ) & = & f_m(\lambda )/f_0(\lambda ). \end{eqnarray}
\reseteqn

Up to second order we will only need the functions $h_m(\lambda )$ for
$m = 1,2,3$, obviously $h_m(\lambda ) = h_{-m}(\lambda )$. The sums in
(4.7) converge rapidly, therefore the consideration of a few terms is
sufficient for the computation.

The evaluation of the full traces is only necessary in the case of
multiply occupied plaquettes. In all configurations where links belong
to one plaquette only, it is much simpler to perform the integrations
over the corresponding link variables first. This removes the
corresponding $\sigma $ - matrices and leads to much simpler traces.
For $m$-fold occupied plaquettes, on the other hand, a calculation by
brute force becomes prohibitively complicated very soon, because
it would involve a sum over $2^{10m}$ terms! Fortunately one can
reduce the complexity of the problem by using a simple group
theoretical relation:

\begin{equation} [\mbox{Tr} U_{1/2}]^2 = 1 + \mbox{Tr} U_1.
\end{equation}
Here $U_{1/2}$ and $U_1$ denote the $SU(2)$ representation matrices
in the spin 1/2 and 1 representation respectively. The latter are
simply related to the former by replacing $e^{i\sigma _k\alpha }$ by
$e^{i2J_k\alpha }$, with $J_k$ the $3\times 3$ representation
matrices. Application of (4.8) reduces the complexity, but not quite
as much as it appears at first sight. The matrices $J_k$ fulfil
$J_k^{2n} = J_k^2$ for $n\geq 1$ and $J_k^{2n+1} = J_k$ for $n \geq
0$. Contrary to the spin 1/2 case, however, $J_k^2\neq 1$. Therefore
the relation corresponding to (4.4) is slightly more complicated:

\begin{equation} e^{i2J_k\alpha } = \sum_{s=0,\pm 1}e^{i2s\alpha }
\bar{P}_s^{(k)},\quad \mbox{with}\quad \bar{P}_0^{(k)} = 1-J_k^2,\quad
\bar{P}_{\pm 1}^{(k)} = \frac{1}{2 }(J_k^2\pm J_k). \end{equation}

Nevertheless, the simplification obtained by using (4.8) is sizeable.
E.g. for the twofold occupied plaquette it reduces the number of
terms in the sum from $2^{20}$ to $3^{10}$, i.e. by a factor of
$\approx 18$. In this way it was possible to calculate all
contributions of the second order without special effort, except the
threefold plaquette. But fortunately the latter appears only once and
can be safely neglected among the 977 configurations which contribute
in second order.

After these remarks we can come to the results:\\[1ex]

{\bf  Order 1 }\\[1ex]

The first order result reads

\begin{eqnarray} \lefteqn{E^{(1)}(\beta ,\lambda ) = }\nonumber\\
& & [1 + \delta \lambda (-2 + 4 h_1'(\lambda )/h_1(\lambda )]
e^{-2\lambda } h_1^4 (\lambda ) \nonumber\\
& & +\delta \beta [(1+e^{-8\lambda }+2 e^{-4\lambda }
h_2^4(\lambda ))/4 - e^{-4\lambda } h_1^8(\lambda)] \nonumber\\
& & +\delta \beta (2d-3) [ \frac{1}{2 } e^{-3\lambda } h_1^6(\lambda )
\{(1+e^{-\lambda })^2 (1 + h_2(\lambda )) +
(1-e^{-\lambda })^2 (1 - h_2(\lambda ))\} \nonumber\\
& & \quad \quad \quad \quad \quad \quad
- 4 e^{-4\lambda } h_1^8(\lambda ) ]. \end{eqnarray}
The first term with $\delta \beta $ arises from the double plaquette,
the second one from the neighboring plaquettes. The whole situation is
rather similar to the U(1) case. The minimum with respect to $\lambda
$ now disappears at $\beta _{pi} = 2.1377$. In fig. 5 we show the
result, together with the one obtained by the power trick and the
(0,1) Pad\'e transformation. The quality of the results is comparable
to the U(1) case. \\[1ex]

{\bf  Order 2 }\\[1ex]

The extremum disappears at $\beta _{pi} = 2.2336$, the corresponding
energy lies somewhat above the data. For increasing $\beta $, however,
the curve drops down below the data as in the U(1) case which shows
again the problematics of even orders of the expansion. We therefore
performed the (1,1) Pad\'e transformation as before. It has an
extremum for all $\beta $ above $\beta _{pi} = 2.2737$. The result is
also shown in fig. 5. The second order Pad\'e transformation reduces
the error by a factor of about one half compared to the first order.
The agreement with the data is not as good as for U(1) in $d=4$. The
poorer quality of the approximation in the non abelian case could be
due to the specific gauge dependence introduced by our definition of
the fields in (4.6), or to the non analyticity in $\vartheta $
resulting from the periodic continuation of the Haar measure in (4.5).

\setcounter{equation}{0}\addtocounter{saveeqn}{1}%

\section{Conclusions}

The present work can be considered as an empirical study of the
convergence properties of the optimized $\delta $ - expansion for
non trivial systems with an infinite number of degrees of freedom and
possibly  a phase transition. Because we were able to go up
to fourth order in the case of U(1) in $d=4$ dimensions, we believe
that our conclusions can be considered as quite reliable. We begin
with the positive aspects:

For $\beta $ above the critical value $\beta _c$ of the phase
transition, one has rapid convergence in the whole region. The
second order (1,1) Pad\'e transformation gives already perfect
agreement with the MC data.

For $\beta <\beta _c$ one needs manipulations like the power trick or
a suitable Pad\'e transformation in order that the principle of
minimal sensitivity can be applied down to lower $\beta $. The
discrepancy with the data is much larger in this region, but a clear
tendency of convergence towards the latter is visible. The fourth
order (2,2) Pad\'e transformation gives a remarkably good
approximation down to $\beta > 0.95$ which hardly can be considered as
accidental. The large increase of the energy within a small region of
$\beta $ is clearly reproduced. This result exceeds the previous work
in refs \cite{DM} -\cite{AJ} where good approximations where obtained
up to the transition region (from above or from below, respectively)
but not beyond. Unfortunately higher order calculations appear not
feasible with reasonable effort without an additional idea. In the
case of U(1) it is quite trivial to write down the contribution for
any configuration with the help of (2.19); the only cumbersome task is
the correct counting of equivalent configurations. Rigorous
convergence proofs for complex systems as considered here are also not
available. So one can only speculate that higher orders would stay
stable above the transition region and further improve the results
below.

Let us finally mention the dubious aspects of the whole approach. Its
``distinctly alchemical flavor'' \cite{Osc} has clearly shown up
again. The ambiguity in the choice of the interpolating action appears
only as a minor deficiency; this choice is an art as in all variation
methods. A really serious problem is that we have no a priori
principle whatever, which tells us which of several extrema should be
chosen. Even worse, there are cases, as in the even orders, where
extrema exist in some $\beta $-interval, but none of them belongs to
the physical situation. There are two possibilities how to find the
relevant extremum, or, alternatively, to reject them all. The first
one is to use additional, at least crude, information, say from MC
data. The second one, which relies completely on the expansion itself,
is to look for the convergence of the solution by comparing different
orders of the expansion. Both criteria can be successfully applied to
our figures. In view of the described general problematics it appears
even more impressive, how the choice of the ``correct'' extremum leads
to excellent results. This should encourage further theoretical work
on the method. \newpage

\setcounter{equation}{0}\addtocounter{saveeqn}{1}%

 \newpage

{\Large \bf Figure Captions}\vspace{2ex}

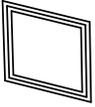
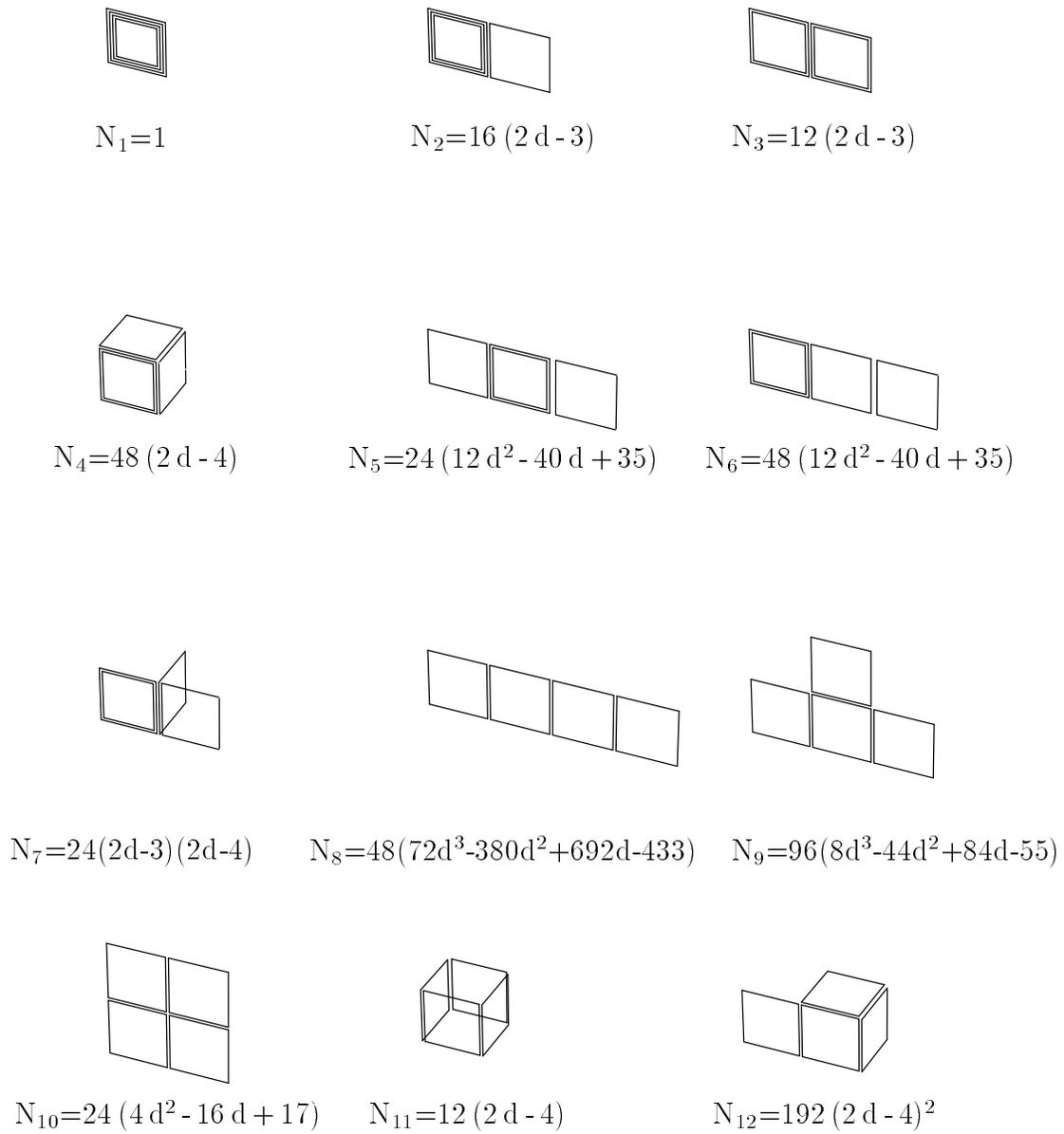
{\bf Fig. 1a,b,c:} The connected plaquette configurations for order 1,
2, and 3\\[1ex]

Together with each generic type we give the number of
configurations of this type for arbitrary dimension $d$. In order 3
the types 8 and 9 give identical contributions, the same holds for
type 10 and 11. One has to note, that, e.g. the symbol for type 8 in
fig. 1c stands for all types of configurations with the same
topology. As an example, some of them are shown in fig. 1d.\\[1ex]

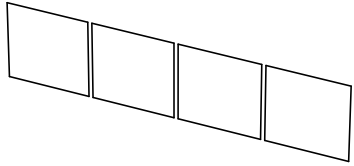
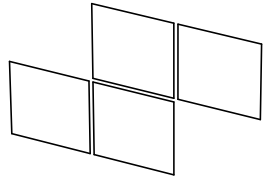
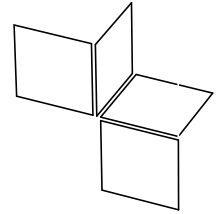
{\bf Fig. 1d:} Some examples for configurations belonging to type 8 of
fig. 1c\\[1ex]

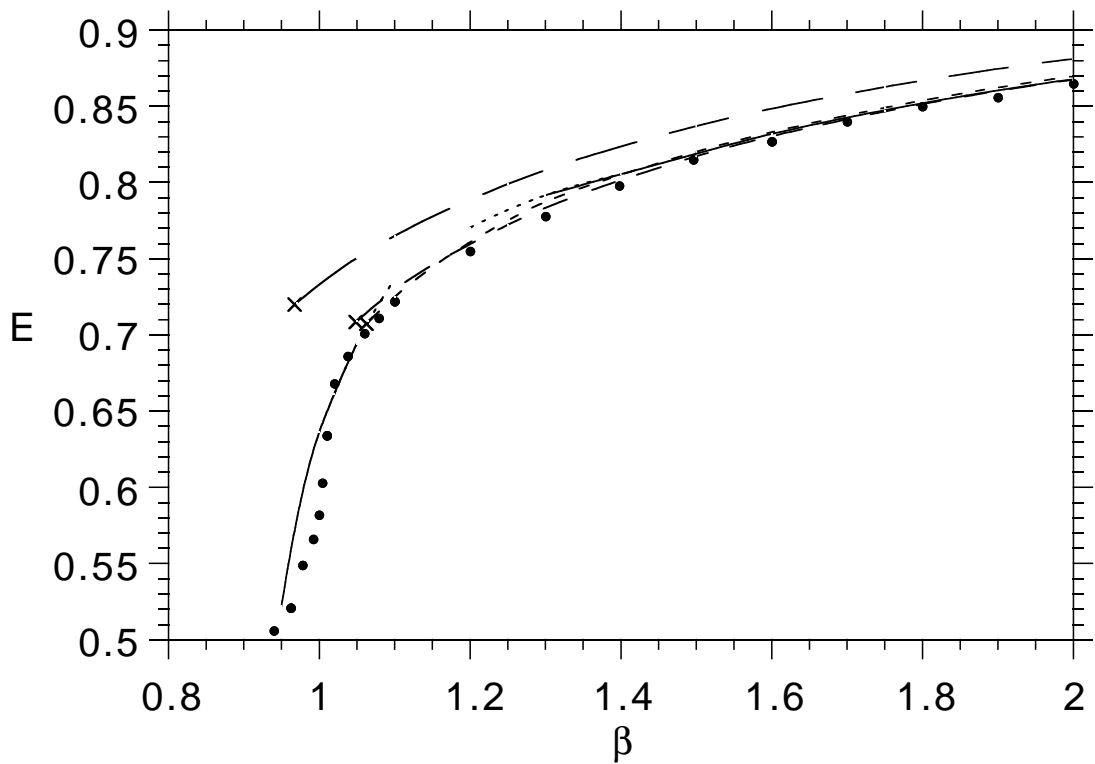
{\bf Fig. 2a:} The U(1) plaquette energy in 4 dimensions near and
above the transition region \\[1ex]

Upper dashed line (long dashes): First order

Dashed line (normal dashes): Second order diagonal Pad\'e
transformation

Dashed line (short dashes): Third order

Solid lines: Fourth order diagonal Pad\'e transformation

The MC data are taken from \cite{Cal}, where also references to
earlier work can be found.  \\[1ex]

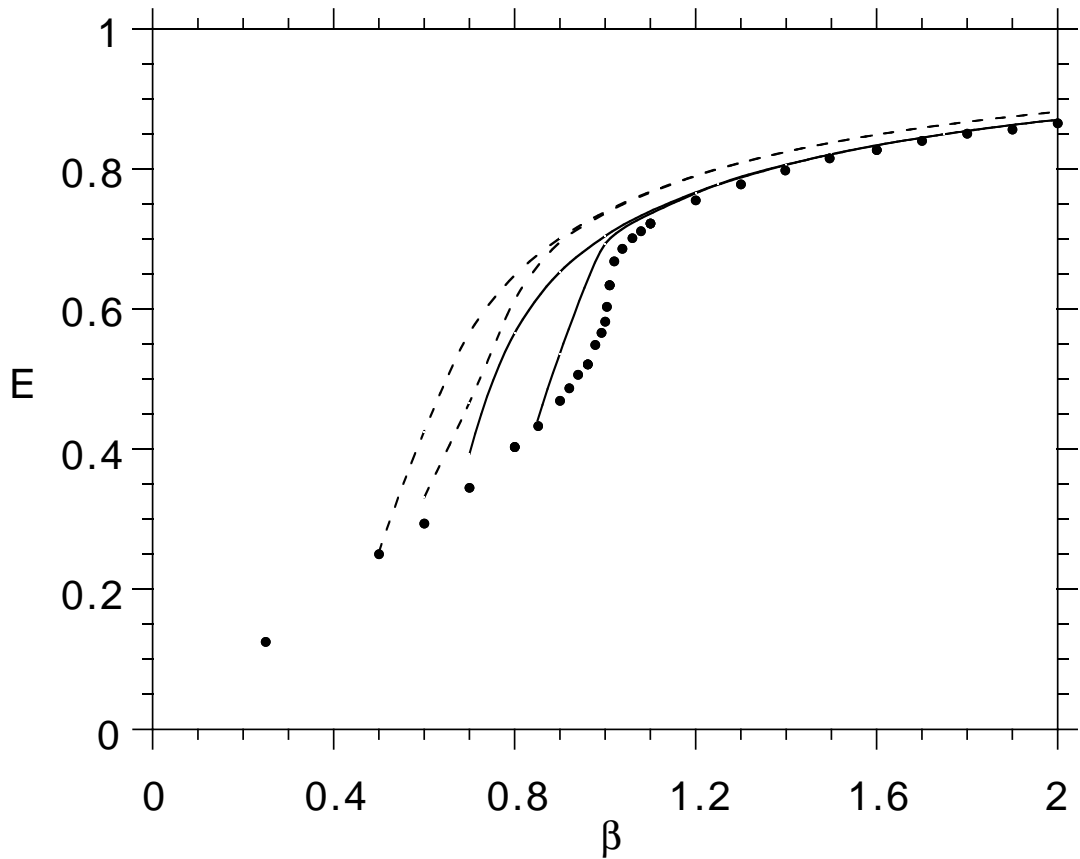
{\bf Fig. 2b:} The U(1) plaquette energy in 4 dimensions extended
below the transition region\\[1ex]

Upper dashed line: First order (0,1) Pad\'e transformation

Lower dashed line: First order power transformation with the
limiting exponent

$\kappa _0 = 0.2781$, for which the continuation to small $\beta $
becomes possible

Upper solid line: Third order (0,3) Pad\'e transformation

Lower solid line: Third order power transformation with limiting
exponent $\kappa _0 = $

0.4182

The MC data are again taken from \cite{Cal}.\\[1ex]

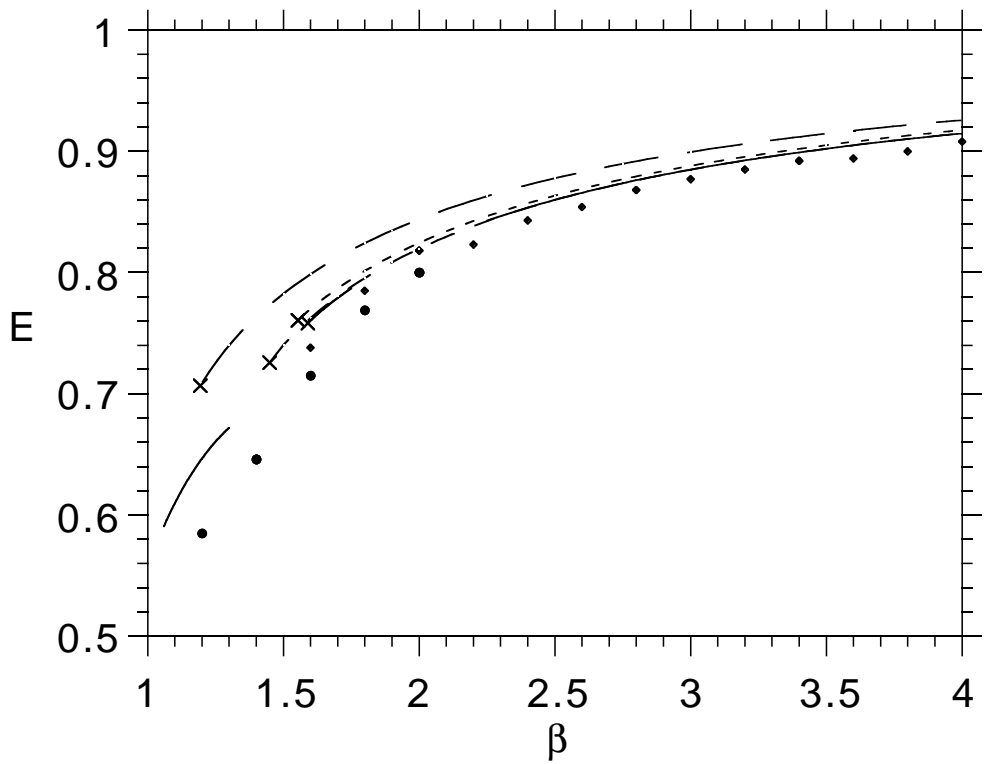
{\bf Fig. 3a:} The U(1) plaquette energy in 3 dimensions near and
above the transition region\\[1ex]

Upper dashed line (long dashes): First order

Dashed line (normal dashes): Second order diagonal Pad\'e
transformation

Dashed line (short dashes): Third order

Solid lines: Fourth order diagonal Pad\'e transformation

The MC data are taken from \cite{BC}.\newpage

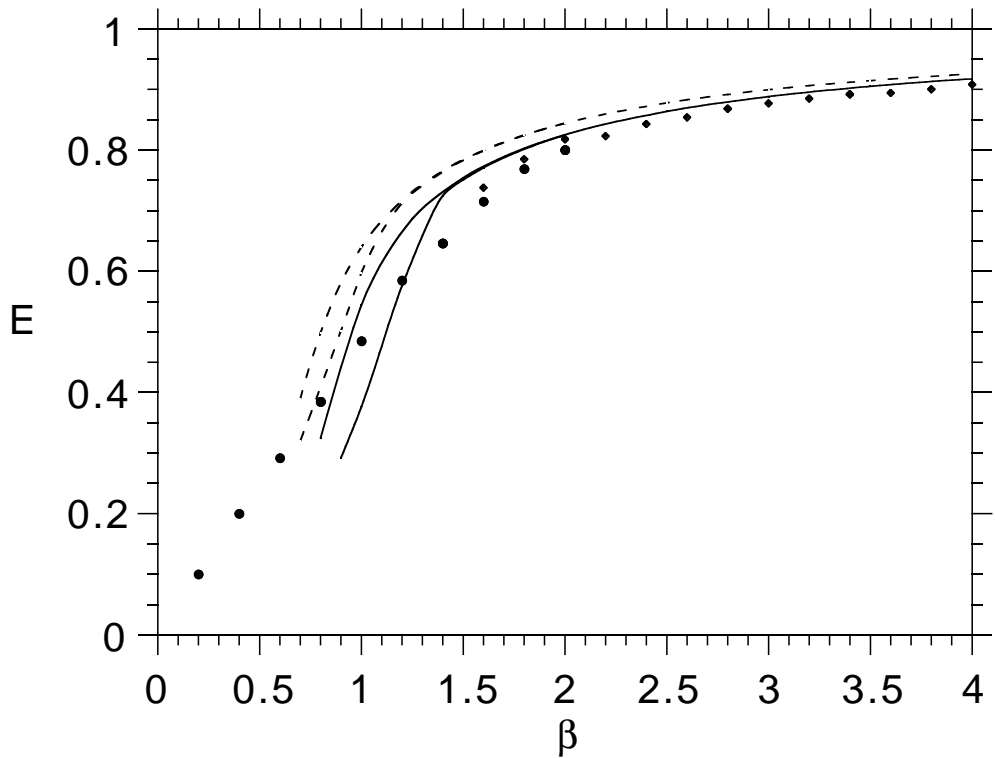
{\bf Fig. 3b:} The U(1) plaquette energy in 3 dimensions extended
below the transition region.\\[1ex]

Upper dashed line: First order (0,1) Pad\'e transformation

Lower dashed line: First order power transformation with  limiting
exponent $\kappa _0 = $

0.5567

Upper solid line: Third order (0,3) Pad\'e transformation

Lower solid line: Third order power transformation with limiting
exponent $\kappa _0 = $

0.5497

The MC data are again taken from \cite{BC}.\\[1ex]

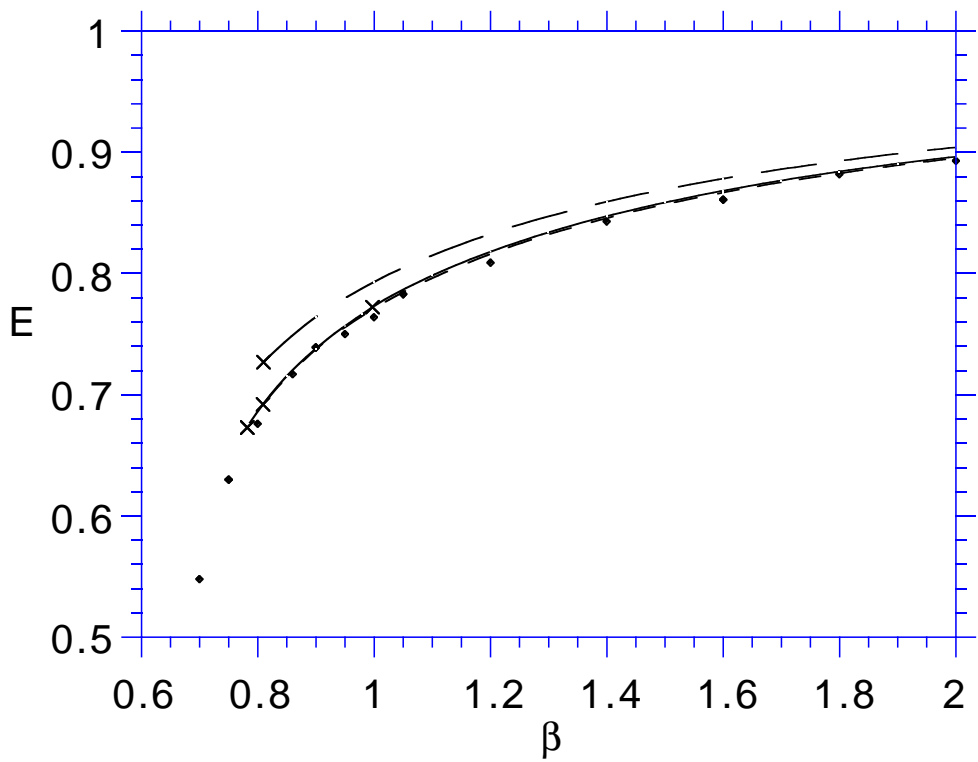
{\bf Fig. 4a:} The U(1) plaquette energy in 5 dimensions near and
above the transition region\\[1ex]

Upper dashed line (long dashes): First order

Dashed line (normal dashes): Second order diagonal Pad\'e
transformation

Solid line: Third order

The MC data are taken from \cite{BC}.\\[1ex]

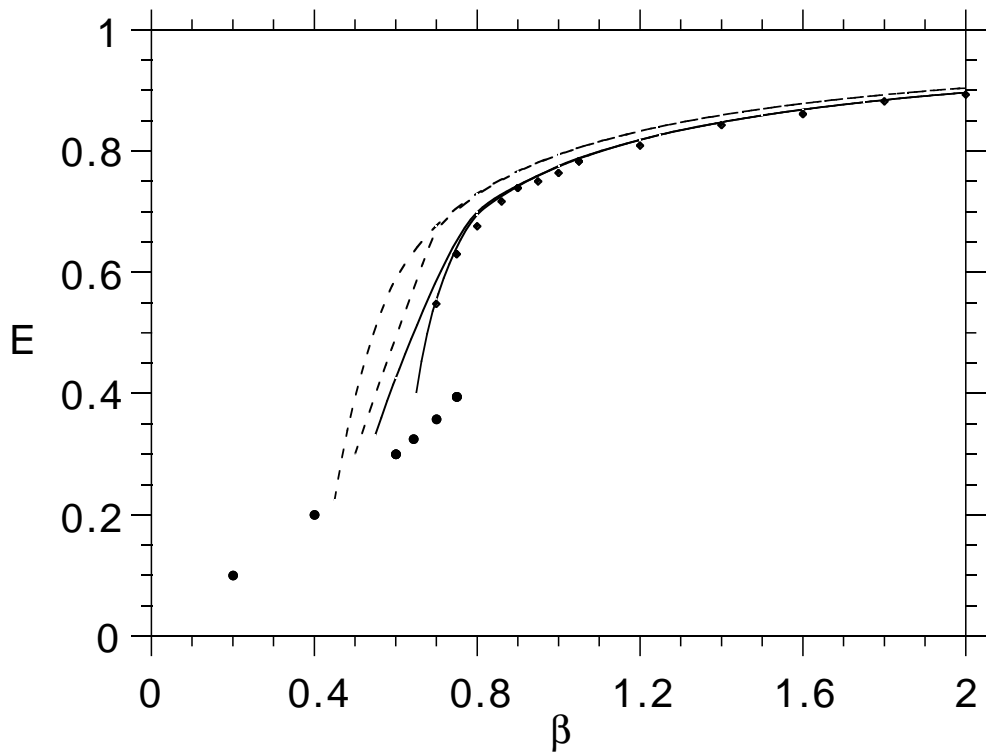
{\bf Fig. 4b:} The U(1) plaquette energy in 5 dimensions extended
below the transition region.\\[1ex]

Upper dashed line: First order (0,1) Pad\'e transformation.

Lower dashed line: First order power transformation with  limiting
exponent $\kappa _0 =$

0.0410

Upper solid line: Third order (0,3) Pad\'e transformation

Lower solid line: Third order power transformation with limiting
exponent $\kappa _0 \approx 0$

The MC data are again taken from \cite{BC}.\\[1ex]

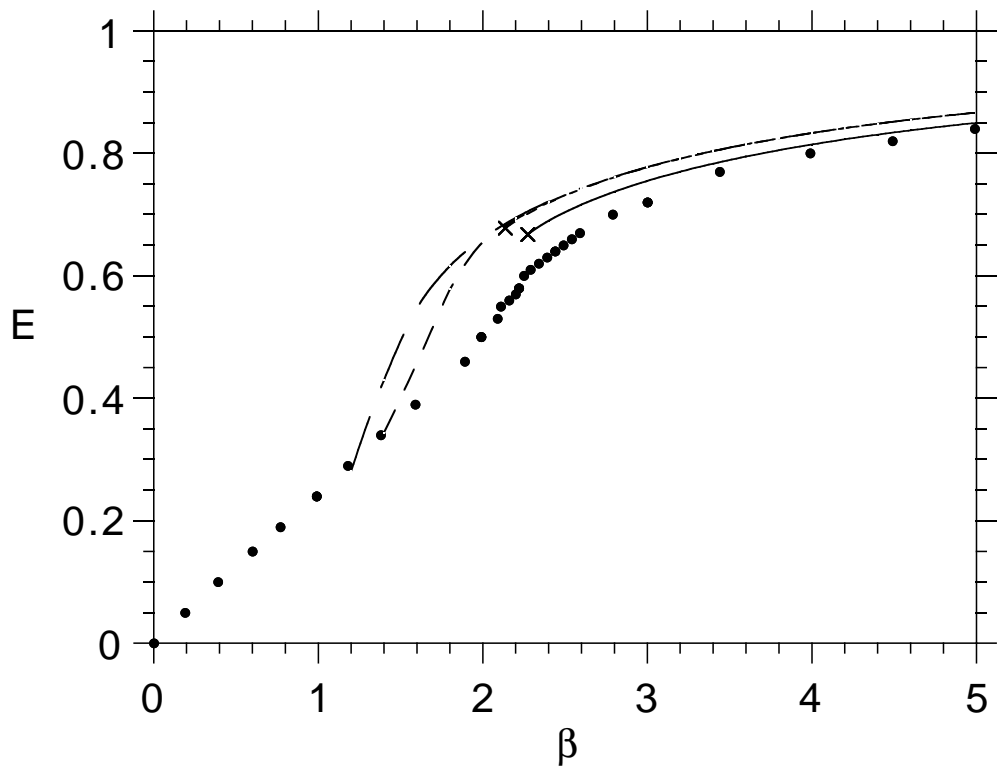
{\bf Fig. 5:} The SU(2) plaquette energy in 4 dimensions.\\[1ex]

Dashed line ending at cross (short dashes): First order

Dashed line (normal dashes): First order (0,1) Pad\'e transformation

Dashed line (long dashes): First order power transformation with
limiting exponent

$\kappa _0 = 0.3968$

Solid line: Second order (1,1) Pad\'e transformation

The MC data are from \cite{LN}.

\end{document}